\documentclass[letterpaper,10pt]{article} 
\usepackage{opticameet3} %% use version 3 for proper copyright statement
\usepackage[nohyperlinks,nolist]{acronym}

\acrodef{AWGN}{additive white Gaussian noise}
\acrodef{BER}{bit error rate}
\acrodef{CMA}{constant modulus algorithm}
\acrodef{CNN}{convolutional neural network}
\acrodef{DSP}{digital signal processing}
\acrodef{EAM}{electro-absorption modulator}
\acrodef{FEC}{forward error correction}
\acrodef{FPGA}{field programmable gate array}
\acrodef{GPU}{graphics processing unit}
\acrodef{MAC}{multiply-accumulate}
\acrodef{ML}{machine learning}
\acrodef{MSE}{mean-squared error}
\acrodef{NN}{neural network}
\acrodef{PCS}{probabilistic constellation shaping}
\acrodef{PON}{passive optical network}
\acrodef{SNN}{spiking neural network}
\acrodef{SOA}{semiconductor optical amplifier}
\acrodef{VAE}{variational auto-encoder}
\acrodef{VQ-VAE}{vector-quantized \ac{VAE}}

\usepackage{float}
\usepackage{wrapfig}
\restylefloat{figure}
\usepackage[font=footnotesize]{caption}
\usepackage{subcaption}
\usepackage{comment}

\usepackage{tikz,mleftright,bm}
\usetikzlibrary{external}
\definecolor{KITgreen}{rgb}{0,.59,.51}
\definecolor{KITblue}{rgb}{.27,.39,.66}
\definecolor{KITred}{rgb}{.63,.13,.13}
\definecolor{KITorange}{rgb}{.87,.60,.10}

\usetikzlibrary{arrows,calc,fit,matrix,positioning,shapes,shadows,trees,mindmap,tikzmark,arrows.meta,angles,quotes,babel}
\tikzstyle{block} = [draw, thick, text=black, align=center, minimum width=1.2cm, minimum height=0.6cm, font=\small]

\newcommand\authormark[1]{\textsuperscript{#1}}

\usepackage[colorlinks=true,bookmarks=false,citecolor=blue,urlcolor=blue]{hyperref} %pdflatex
\usepackage{siunitx}
\usepackage{pgfplots}
\pgfplotsset{compat=newest} 

\newcommand\blfootnote[1]{%
  \begingroup
  \renewcommand\thefootnote{}\footnote{#1}%
  \addtocounter{footnote}{-1}%
  \endgroup
}

\begin{document}

\title{Recent Advances on Machine Learning-aided DSP for Short-reach and Long-haul Optical Communications}

\author{Laurent Schmalen,\authormark{1,*} Vincent Lauinger,\authormark{1} Jonas Ney,\authormark{3} Norbert Wehn,\authormark{3} Patrick Matalla,\authormark{2} Sebastian Randel,\authormark{2} Alexander von Bank,\authormark{1} and Eike-Manuel Edelmann\authormark{1}}

\address{\authormark{1}Communications Engineering Lab (CEL), Karlsruhe Institute of Technology (KIT), 76131 Karlsruhe, Germany\\ 
\authormark{2}Institute of Photonics and Quantum Electronics (IPQ), Karlsruhe Institute of Technology (KIT), 76131 Karlsruhe, Germany\\
\authormark{3}Microelectronic Systems Design (EMS), RPTU Kaiserslautern-Landau, 67653 Kaiserslautern, Germany} %\hspace{10ex}\authormark{*}vincent.lauinger@kit.edu}

\email{\authormark{*}\texttt{laurent.schmalen@kit.edu}} %% email address is required

\begin{abstract}
In this paper, we highlight recent advances in the use of machine learning for implementing equalizers for optical communications. We highlight both algorithmic advances as well as implementation aspects using conventional and neuromorphic hardware.
\end{abstract}

\section{Introduction}\vspace{-1ex}
In the past decade, the field of \ac{ML} has seen a tremendous spike in popularity and has led to transformative changes in almost every field of science and engineering. This is mostly due to the success of \acp{NN} and in particular the technique of deep learning~\cite{lecun2015deep}. Deep learning and
the accompanying software tools have also found their way into optical communications and are now indispensable tools in the field; ML is now commonly used in all parts of fiber-optical communication networks~\cite{Agrell_2024}.\blfootnote{This work has received funding from the European Research Council (ERC) under the European Union’s Horizon 2020 research and innovation program (grant agreement No. 101001899). Parts of this work were carried out in the framework of the CELTIC-NEXT project AI-NET-ANTILLAS (C2019/3-3, grant 16KIS1316 and 16KIS1317), funded by the German Federal Ministry of Education and Research (BMBF). The authors thank Jinxiang Song from Chalmers University of Technology for valuable discussions. }

ML is already widely used for parameter estimation in optical networks, with the goal of configuring optical network links. Due to their capacity as universal function approximators, ML algorithms and in particular \acp{NN} are also often used in the physical layer to replace suboptimal or overly complex \ac{DSP} algorithms in the receiver or transmitter. 

The use of ML to replace parts of the transmitter or receiver,
e.g., as \ac{DSP} algorithms or to support \ac{FEC} decoding, still poses many research challenges, despite the benefits we already see. In particular, standard out-of-the-box \ac{ML} solutions typically have higher computational complexity than conventional, optimized algorithms. Due to the enormous data rates at which optical communication systems operate, complexity is a major concern. The parallel structure of \acp{NN} can lead to straightforward parallelization (as in the ubiquitous \ac{GPU} implementations), which makes them attractive for implementation in optical transceivers. A future challenge will be the development of ultra-low-complexity hardware platforms with low power dissipation that can be used in
highly integrated, high-speed optical transceivers. 

In this paper, we present recent advances in the use of ML-based equalizers for optical communications. We focus on two scenarios: coherent long-haul communications and short-reach intensity-modulation/direct-detection (IM/DD) systems. We demonstrate how the theory developed in the \ac{ML} field can be beneficially used to derive new equalizers specifically tailored to optical communications. Furthermore, we explore how \ac{ML}-based equalizers for short-reach applications can be efficiently implemented, both on traditional digital hardware and on neuromorphic hardware. We would like to emphasize that the use of \ac{ML} to optimize \ac{DSP} algorithms is not restricted to equalization, but can be expanded to other parts of the \ac{DSP} chain, e.g., phase estimation~\cite{Rode23JLT}.

\section{Novel ML-inspired Cost Functions for Conventional Equalizers}\vspace{-1ex}
The use of ML as an equalizer is natural, as ML-based equalizers can learn to undo the effects of a channel based on training data. If pilot symbols are available, they can be used with \emph{supervised learning} to train  an \ac{ML} model (typically an \ac{NN} or \ac{CNN}) that undoes the effect of the channel. However, when using the typical \ac{MSE} as a cost function, the equalized constellation can exhibit the so-called ``jail-window'' effect~\cite{freire2022neural}. Such a resulting constellation can have negative effects on the demapping and consequently on \ac{FEC} decoding. By adjusting the cost function or by taking into account the effect of the demapper into the cost function, the jail-window effect can be prevented~\cite{diedolo2022nonlinear}. 

\begin{figure} [tb]
		\centering
       \includegraphics{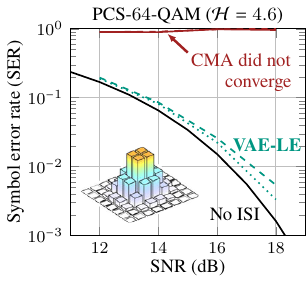}
        \hspace*{1.2ex}
        \includegraphics{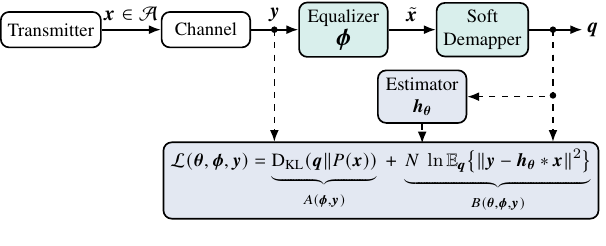}
		\vspace*{-1.5ex}
		\caption{Left: simulation results of the VAE-LE and the CMA for the transmission of 64-QAM with probabilistic constellation shaping (PCS) (entropy $\mathcal{H}=4.6$) over a linear dispersive optical dual-polarization channel at a symbol rate of $40 \ \mathrm{GBd}$ (dotted) and $90 \ \mathrm{GBd}$ (dashed)~\cite{lauinger2022blind}. Right: block diagram of the VAE based equalizer.}
		\label{fig:vae}
		%\vspace*{-4ex}
\end{figure}

The idea of jointly considering equalizer and demapper opens further new possibilities, which are explored in~\cite{lauinger2022blind}. The use of a \ac{VAE} yields a novel \emph{blind} cost function for an equalizer, i.e., to adapt the equalizer, we only require the received, noisy symbols (as well as the constellation and source symbol statistics). In contrast to traditional cost functions (e.g., the \ac{CMA}), the output of the demapper is used to adjust the equalizer taps. This cost function can be used both with traditional linear equalizers (denoted VAE-LE), but also with novel \ac{NN}- or \ac{CNN}-based equalizers~\cite{caciularu2020unsupervised}. We have furthermore shown in~\cite{lauinger2022blind} that these equalizers can achieve outstanding performance if \ac{PCS} is used---a scenario where classical equalizers, e.g., the \ac{CMA} fail. Figure~\ref{fig:vae} shows the general block diagram of \ac{VAE}-based equalizers and simulation results for a long-haul transmission with \ac{PCS}~\cite{lauinger2022blind}, outperforming a \ac{CMA}-based equalizer. \ac{VAE}-based equalizers furthermore offer advantages in the startup phase, i.e., during initial locking~\cite{Lauinger23OFC}. The \ac{VAE} also includes a channel estimator, which directly yields an estimate of the (possibly time-varying) impulse response of the channel, which can be used, e.g., for integrated communications and sensing~\cite{lauinger2022blind}.

Unfortunately, the mathematical derivation of the novel \ac{VAE}-based cost function relies on the assumption of a linear channel and \ac{AWGN}. In many scenarios in optical communications, e.g., when nonlinearities due to the Kerr effect are present, we can still use the benefits of the \ac{VAE} through a simplification called \ac{VQ-VAE}, which offers simple blind equalization when the implementation of the \ac{VAE} is more challenging~\cite{song2023blind,Song23ECOC}.

 %\vspace*{-10.5ex}

\begin{figure} [tb]
    \centering
     \includegraphics{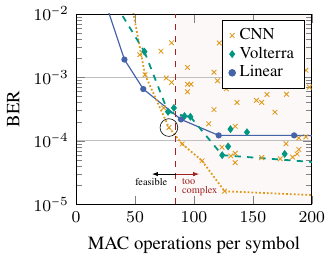}
     \includegraphics{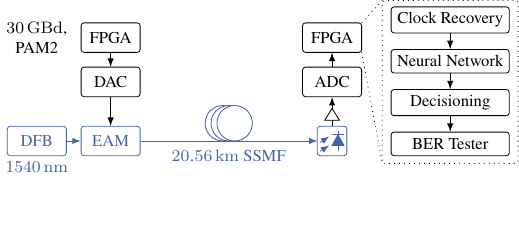}
    \vspace*{-1ex}
    \caption{Left: exploration of the performance complexity trade-off of CNN, Volterra, and linear equalizers for IM/DD transmission with pulse-amplitude modulation (PAM)-2 at $40 \ \mathrm{GBd}$. The red dashed vertical line approximates the complexity limit on the AMD XCVU13P FPGA~\cite{neySAMOSJournal}. The lines connecting the points give the Pareto fronts and the circle marks the Pareto optimal model for this system. Right: Transmission setup of an FPGA implementation demonstrating real-time ML-based equalization~\cite{Ney24ICMLCN}.\vspace*{-2ex}}
    \label{fig:fpga_impl}
\end{figure}

\section{Novel ML-based Equalizers and Their Hardware Implementation}\vspace{-1ex} 
It is a common misconception that ML-based equalizers are more computationally demanding than conventional (linear or nonlinear) equalizers. This misconception mostly stems from the large number of multiplications in typical fully-connected \ac{NN} and typical \ac{CNN} layers. However, we can show that with a careful optimization of the structure of the \ac{CNN} (i.e., the number of layers, layer configuration, number of neurons per layer, etc.), we can outperform linear and nonlinear Volterra equalizers in a short-reach IM/DD scenario, if the number of operations is constrained. The constraint on operations was due to the real-time implementation of the equalizer on a \ac{FPGA}~\cite{Ney23SAMOS,Ney24ICMLCN}. We have successfully tested and evaluated such an equalizer for high-speed \acp{PON}, where nonlinear effects caused by low-cost components such as the gain saturation in \acp{SOA} or the nonlinear transfer functions of \acp{EAM} strongly distort the signal nonlinearly in addition to chromatic dispersion~\cite{Lauinger24OFC}.

Figure~\ref{fig:fpga_impl} shows the design space exploration for a short-reach IM/DD system, where we compare the performance of a linear equalizer, a nonlinear Volterra equalizer and a \ac{CNN}-based equalizer (with different configurations) as a function of the required \ac{MAC} operations per symbol. With a given \ac{FPGA} platform, which constrains the number of \ac{MAC} operations for a baud rate of $40$\,GBd, we observe that one instance of the \ac{CNN} equalizer achieves the lowest \ac{BER}. Furthermore, we demonstrated that even an ML-based equalizer including the real-time training can be efficiently implemented on an FPGA up to a data rate of $20$\,GBd~\cite{ney2024training}.

\section{Energy-efficient Equalizers with Spiking Neural networks}\vspace{-1ex}
A promising research direction to further improve the energy efficiency of equalizers for optical communications is the use of neuromorphic hardware. Various neuromorphic hardware architectures exist; an interesting biology-inspired architecture is \acp{SNN}, which we have used for an energy-efficient electronic implementation of equalizers. \acp{SNN} are discrete, event-based variants 
\begin{wrapfigure}{r}{0.5\textwidth}
\includegraphics[width=0.5\textwidth]{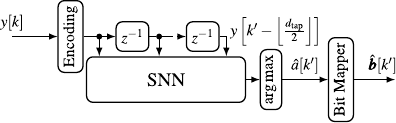}
\caption{Feed-forward SNN-based equalizer for short-reach optical communications according to~\cite{Bank24OFC}.}\label{fig:snn}
\end{wrapfigure}
of \acp{NN} inspired by the biological signal processing in the brain. \acp{SNN} have advantages in terms of circuit power consumption, as energy is only used sporadically to emit so-called spikes. \acp{SNN} have been used successfully as equalizers in short-reach optical communications~\cite{arnold2023spiking,Bansbach23WSASCC,Bank24OFC} (with one structure being shown in Fig.~\ref{fig:snn}). Recently, it was shown that even a digital emulation of such an \ac{SNN}-based equalizer can have advantages in terms of energy efficiency compared to a classical digital equalizer~\cite{moursi2024efficient}.

\section{Conclusion}\vspace{-1ex}
In this paper, we have highlighted that \ac{ML} can open many interesting research perspectives in the implementation of \ac{DSP} algorithms for short-reach and long-haul optical communications, from algorithm design to energy-efficient implementation.
\vspace*{1ex}

\footnotesize 


\begin{thebibliography}{10}
\providecommand{\url}[1]{#1}
\csname url@samestyle\endcsname
\providecommand{\newblock}{\relax}
\providecommand{\bibinfo}[2]{#2}
\providecommand{\BIBentrySTDinterwordspacing}{\spaceskip=0pt\relax}
\providecommand{\BIBentryALTinterwordstretchfactor}{4}
\providecommand{\BIBentryALTinterwordspacing}{\spaceskip=\fontdimen2\font plus
\BIBentryALTinterwordstretchfactor\fontdimen3\font minus
  \fontdimen4\font\relax}
\providecommand{\BIBforeignlanguage}[2]{{%
\expandafter\ifx\csname l@#1\endcsname\relax
\typeout{** WARNING: IEEEtran.bst: No hyphenation pattern has been}%
\typeout{** loaded for the language `#1'. Using the pattern for}%
\typeout{** the default language instead.}%
\else
\language=\csname l@#1\endcsname
\fi
#2}}
\providecommand{\BIBdecl}{\relax}
\BIBdecl

\bibitem{lecun2015deep}
Y.~LeCun, Y.~Bengio, and G.~Hinton, ``Deep learning,'' \emph{Nature}, vol. 521,
  no. 7553, pp. 436--444, 2015.

\bibitem{Agrell_2024}
\BIBentryALTinterwordspacing
E.~Agrell \emph{et~al.}, ``Roadmap on optical communications,'' \emph{Journal
  of Optics}, vol.~26, no.~9, p. 093001, Jul. 2024. [Online]. Available:
  \url{https://dx.doi.org/10.1088/2040-8986/ad261f}
\BIBentrySTDinterwordspacing

\bibitem{Rode23JLT}
A.~Rode, B.~Geiger, S.~Chimmalgi, and L.~Schmalen, ``End-to-end optimization of
  constellation shaping for {Wiener} phase noise channels with a differentiable
  blind phase search,'' \emph{J. Lightw. Technol.}, vol.~41, no.~12, pp.
  3849--3859, Jun. 2023, \url{https://arxiv.org/abs/2212.03839}.

\bibitem{freire2022neural}
P.~J. Freire \emph{et~al.}, ``Neural networks-based equalizers for coherent
  optical transmission: Caveats and pitfalls,'' \emph{{IEEE} J. Sel. Topics
  Quantum Electron.}, vol.~28, no. 4: Mach. Learn. in Photon. Commun. and Meas.
  Syst., pp. 1--23, 2022.

\bibitem{diedolo2022nonlinear}
F.~Diedolo, G.~B{\"o}cherer, M.~Sch{\"a}dler, and S.~Calabr{\'o}, ``Nonlinear
  equalization for optical communications based on entropy-regularized mean
  square error,'' in \emph{Proc. ECOC}, 2022, pp. We2C--2.

\bibitem{lauinger2022blind}
V.~Lauinger, F.~Buchali, and L.~Schmalen, ``Blind equalization and channel
  estimation in coherent optical communications using variational
  autoencoders,'' \emph{{IEEE} J. Sel. Areas Commun.}, vol.~40, no.~9, pp.
  2529--2539, 2022, \url{https://arxiv.org/abs/2204.11776}.

\bibitem{caciularu2020unsupervised}
A.~Caciularu and D.~Burshtein, ``Unsupervised linear and nonlinear channel
  equalization and decoding using variational autoencoders,'' \emph{IEEE Trans.
  on Cogn. Commun. Netw.}, vol.~6, no.~3, pp. 1003--1018, 2020.

\bibitem{Lauinger23OFC}
V.~Lauinger, F.~Buchali, and L.~Schmalen, ``Improving the bootstrap of blind
  equalizers with variational autoencoders,'' in \emph{Proc. Opt. Fiber Commun.
  Conf. (OFC)}, San Diego, CA, USA, Mar. 2023,
  \url{https://arxiv.org/abs/2301.06576}.

\bibitem{song2023blind}
J.~Song \emph{et~al.}, ``Blind channel equalization using vector-quantized
  variational autoencoders,'' \emph{preprint}, 2023,
  \url{https://arxiv.org/abs/2302.11687}.

\bibitem{Song23ECOC}
------, ``Blind frequency-domain equalization using vector-quantized
  variational autoencoders,'' in \emph{Proc. ECOC}, Glasgow, UK, Oct. 2023,
  \url{https://arxiv.org/abs/2312.16003}.

\bibitem{neySAMOSJournal}
J.~Ney \emph{et~al.}, ``{CNN}-based equalization for communications: Achieving
  gigabit throughput with a flexible {FPGA} hardware architecture,''
  \emph{\emph{submitted to} International Journal of Parallel Programming},
  2024, \url{https://arxiv.org/abs/2405.02323}.

\bibitem{Ney24ICMLCN}
------, ``Real-time {FPGA} demonstrator of {ANN}-based equalization for optical
  communications,'' in \emph{Proc. ICMLCN}, Stockholm, Sweden, May 2024, demo
  session, \url{https://arxiv.org/abs/2402.15288}.

\bibitem{Ney23SAMOS}
------, ``From algorithm to implementation: enabling high-throughput
  {CNN}-based equalization on {FPGA} for optical communications,'' in
  \emph{Proc. SAMOS XXIII}, Samos, Greece, Jul. 2023.

\bibitem{Lauinger24OFC}
V.~Lauinger \emph{et~al.}, ``Fully-blind neural network based equalization for
  severe nonlinear distortions in 112 {Gbit/s} passive optical networks,'' in
  \emph{Proc. OFC}, San Diego, CA, USA, Mar. 2024,
  \url{https://arxiv.org/abs/2401.09579}.

\bibitem{ney2024training}
J.~Ney and N.~Wehn, ``Achieving high throughput with a trainable
  neural-network-based equalizer for communications on {FPGA},'' in \emph{Proc.
  DSD}, Aug. 2024, \url{https://arxiv.org/abs/2407.02967}.

\bibitem{Bank24OFC}
A.~von Bank, E.-M. Edelmann, and L.~Schmalen, ``Energy-efficient spiking neural
  network equalization for {IM/DD} systems with optimized neural encoding,'' in
  \emph{Proc. Opt. Fiber Commun. Conf. (OFC)}, San Diego, CA, USA, Mar. 2024,
  \url{https://arxiv.org/abs/2312.12909}.

\bibitem{arnold2023spiking}
E.~Arnold \emph{et~al.}, ``Spiking neural network nonlinear demapping on
  neuromorphic hardware for {IM/DD} optical communication,'' \emph{J. Lightw.
  Technol.}, vol.~41, no.~11, pp. 3424--3431, 2023.

\bibitem{Bansbach23WSASCC}
E.-M. Bansbach, A.~von Bank, and L.~Schmalen, ``Spiking neural network decision
  feedback equalization,'' in \emph{Proc. WSA-SCC}, Braunschweig, Germany, Feb.
  2023, \url{https://arxiv.org/abs/2211.04756}.

\bibitem{moursi2024efficient}
M.~Moursi, J.~Ney, B.~Hammoud, and N.~Wehn, ``Efficient {FPGA} implementation
  of an optimized {SNN}-based {DFE} for optical communications,'' in
  \emph{Proc. IEEE MECOM}, Nov. 2024, \url{https://arxiv.org/abs/2409.08698}.

\end{thebibliography}
\end{document}